\documentclass[a4paper,11pt]{article}
\usepackage{amsmath}
\usepackage{amsfonts}
\usepackage[mathscr]{euscript}
\usepackage{bm}
\usepackage{graphicx}
\usepackage{subfigure}
\usepackage{jheppub}
\usepackage[T1]{fontenc}
\usepackage{ulem}
\newcommand{\bea}{\begin{eqnarray}}
\newcommand{\eea}{\end{eqnarray}}
\newcommand{\beq}{\begin{equation}}
\newcommand{\eeq}{\end{equation}}

\def\/{\over}

\title{Thermodynamics of charged accelerating  AdS black holes and holographic heat engines }
\author{Jialin Zhang, Yanjun Li,  and Hongwei Yu\footnote{Corresponding author at hwyu@hunnu.edu.cn} }

\affiliation {Department of Physics and Synergetic Innovation Center for Quantum Effects and Applications, Hunan Normal University, Changsha, Hunan 410081, China}

\abstract{
Using a reasonable choice in normalizing the timelike Killing vector, we  investigate  the thermodynamic properties of
charged accelerating Anti-de Sitter (AdS) black holes.  We find that the  expression of
 the thermodynamic mass in the first law of thermodynamics displays an inextricably intertwining behavior with the charge due to the
unusual  asymptotic structure of the accelerating black holes. Meanwhile, the thermodynamic
length as a potential conjugate to the varying cosmic string
tension is introduced and analyzed in detail, and the possible phase behavior of the charged accelerating black holes is also discussed  in a  standard thermodynamic analysis. Furthermore, we
also investigate the properties of holographic heat engines with
charged accelerating AdS black holes as the working substance in a
benchmarking scheme. We find that the efficiencies of the black hole
heat engines can be influenced by both the size of the benchmark
circular cycle and the cosmic string tension.  More interestingly,
 the existence of charge may significantly increase the
efficiencies of the black hole heat engines and make them be more
sensitive to a varying cosmic string tension. A cross-comparison of
the efficiencies of different black hole heat engines suggests that
the acceleration also increases the efficiency and renders it more
sensitive as the charge varies.}


\begin{document}
\maketitle
\flushbottom
\section{Introduction}
Ever since Hawking's  pioneering work on the radiation of
black holes~\cite{S.H 1975,S.H 1976}, the  black hole thermodynamics
has become an important topic of intensive research and been considered as
a key to gaining insight into the quantum nature
of gravity.
In the last decades, the thermodynamics of several classes of black
holes has been studied, and remarkable progresses have been made in, for example,
 connecting the laws of gravitation with the laws of
thermodynamics~\cite{Jacobson:1995,Padmanabhan:2010}, deploying the
black holes as a  holographically dual system in quantum
chromodynamics~\cite{Kovtun:2005} and condensed matter
physics~\cite{Hartonll:2007,Hartnoll:2008},  and approaching the
thermodynamics of black holes
geometrically~\cite{Aman:2003,Quevedo:2008,Ruppeiner:2014,Mansoori:2014,Suresh:2014,Zhang:2015,Zhang-2:2015,Hendi:2015,Quevedo:2016,
Banerjee:2017}.

Recent studies show that  the thermodynamical properties of black holes in anti-de Sitter (AdS) space are quite different from those in flat
or de Sitter(dS) space. In particular, a subject dubbed
black hole chemistry ~\cite{Kubiznak:2015,Mann:2016}, in which  the negative
cosmologic constant $\Lambda$ is regarded as a thermodynamic
pressure in the extended phase
space~\cite{Teitelboim:1985,Caldarelli:2000,Kastor:2009,Dolan:2011,Dolan:2011-2,Cvetic:2011,Kubiznak:2017},
has started to attract  a growing deal of interest. To date, the investigation in  black hole chemistry has led to the discovery of  a lot of
interesting thermodynamic phenomena, such as the analogous phase
of van der Waals fluid~\cite{Kubiznak:2012}, re-entrant phase
transitions~\cite{Altamirano:2013} and the analogous superfluid phase
transitions~\cite{Hennigar:2017}.

In the context of black hole chemistry, the concept of
holographic heat engines which can extract work with the AdS black holes as the working
substance in cycles in the pressure-volume
phase plane was proposed by Johnson
~\cite{Johnson:2014}.  Subsequently, in order to better crossly compare the
efficiency of the heat engines with different black holes as the working
material, a circular cycle of the heat engine in the pressure-volume
plane has been introduced as a benchmarking prescription by
Chakraborty and Johnson~\cite{Chakraborty:2016}.
After Johnson's pioneering work, the concept of heat engines  has been generalized
 to other different black holes (for an
incomplete list of references,
see~\cite{Belhaj:2015,Caceres:2015,setare:2015,Johnson:2016-1,Johnson:2016-2,Zhang:2016,Sadeghi:2017,Wei:2017,Hendi:2017,Bhamidipati:2017,Xu:2017,Liu:2017,Mo:2017,Hennigar:2017-1,Zhang:2018,Panah:2018}).
Particularly, the properties of the holographic heat engine with
accelerating AdS black holes as the working substance have drawn
considerable interest. The accelerating AdS black holes are
described by the so-called
$C-$metric~\cite{Kinnersley:1970,Plebanski:1976,Dias:2003,Griffiths:2006},
which can be used to study either one or two accelerating black
holes, the pair creation of black holes~\cite{Dowker:1994}, the
construction  of the black ring solution in 5-dimensional gravity
~\cite{Emparan:2002}, and the test of gravitational
radiation~\cite{Podolsky:2003}.

However,  the $C-$metric, despite of the aforementioned various
applications,  remains somewhat esoteric in its thermodynamic
prescription  partly due to the fact that it has an unusual
asymptotic behavior and  at least one non-removable conical
singularity on the azimuthal axis of symmetry. As a result,
discrepant and conflicting results have been obtained in the
investigation of the thermodynamics of the accelerating AdS black
holes~\cite{Appels:2016,Appels:2017,Gregory:2017,Astorino:2017,Anabalon:2018}.
For example, a set of thermodynamic variables was obtained in
Refs.~\cite{Appels:2016,Appels:2017,Gregory:2017}, where the
 thermodynamic mass and temperature are associated with
the coordinate time rather than with the normalized timelike Killing
vector. However, the corresponding first law of
thermodynamics was later found to be incorrect~\cite{Appelsphd:2018}.

 Let us  note that
alternate expressions for the thermodynamic mass
has also been posited by assuming the integrability condition in Ref.~\cite{Astorino:2017}. It has been argued that for regular accelerating black holes, the acceleration is provided by an external electromagnetic field rather than the conical deficit associated with a string, while for the irregular $C$-metric with conical singularities, the black hole is  again accelerated by a pulling string.  However, in latter case, the cosmic string tension, which should be considered as a thermodynamic
parameter relevant to the varying conical deficit,  was not included
in the first law.

Recently, A. Anabalon et al
re-examined the thermodynamics of uncharged accelerating AdS black
holes by means of the mass and temperature generated by a normalized
time that just corresponds to the ``time" of an asymptotic
observer~\cite{Anabalon:2018}. Such generated thermodynamic mass has
been verified to be consistent with the  mass obtained by
holographic computing, and hence it  reconciles discrepancies and conflicts  in the previous investigations and confirms at this point that
such choice of normalization of timelike Killing vector is
applicable and well grounded.  Based on the correct thermodynamics  of uncharged
accelerating AdS black holes stated in Ref.~\cite{Anabalon:2018},
the properties of the corresponding benchmarking holographic heat
engines have been studied in detail~\cite{Zhang:2018}. However, a formulation of the first law of thermodynamics and an examination of the holographic heat engines when the
accelerating black hole is charged, where the additional special asymptotic structure
of the gauge field renders the thermodynamics a non-trivial task,  are quite challenging~\cite{Anabalon:2018}. More recently, A. Anabalon et al pointed out in Ref.~\cite{Anabalon:2018-2}  that the choice of  normalizing timelike Killing vector can be appropriately fixed by holographic methods for charged accelerating black holes, which takes into
account of that  all  the thermodynamic parameters are analogous to those of an observer who is comoving with an accelerating charged object and does not see any radiation.  The purpose of the present paper is  to explore the consistent thermodynamic properties  for the charged accelerating AdS black holes and study
the influence of charge on the corresponding holographic heat engines.

This paper is organized as follows.
We begin in section~\ref{II} by  exploring the explicit expressions for the
thermodynamic parameters of charged accelerating AdS black holes
with an appropriate normalized time. In section~\ref{III}, we
investigate the properties of the benchmarking holographic heat
engines with charged accelerated AdS black holes as working
substances. We  end with a summary and a conclusion in section~\ref{IV}.

\section{The Charged accelerating AdS black hole  and its thermodynamics }
\label{II}
 A charged accelerating AdS black hole is described by the following  $C-$metric \cite{Griffiths:2006,Appels:2016,Appels:2017}
\begin{align}\label{c-metric}
ds^{2}=\frac{1}{\Omega^{2}}\bigg[f(r)dt^{2}-\frac{dr^{2}}{f(r)}-r^{2}\Big(\frac{d\theta^{2}}{g(\theta)}
+g(\theta)\sin^{2}\theta\frac{d\phi^{2}}{K^{2}}\Big)\bigg]\;,
\end{align}
where
\begin{equation}
\Omega=1+A r\cos\theta\;,
\end{equation}
\begin{equation}\label{fr}
f(r)=(1-A^{2}r^{2})\Big(1-\frac{2m}{r}+\frac{e^{2}}{r^{2}}\Big)+\frac{r^{2}}{\ell^{2}}\;,
\end{equation}
\begin{equation}
g(\theta)=1+2mA \cos\theta+e^{2}A^{2}\cos^{2}\theta\;.
\end{equation}
The corresponding gauge potential can be considered to be~\cite{Anabalon:2018-2}
\begin{equation}\quad B=-\frac{e}{r}dt+\frac{e}{r_+}dt\;\end{equation}
with $r_{+}$ representing the black hole event horizon. It is easy to deduce that this gauge potential vanishes at the horizon.
 Here, the parameters $m$, $e$ and  $A$ are related to
the mass, the electric charge and  the magnitude of acceleration of
the black hole respectively, $\ell$ represents the AdS radius, and
$K$ characterizes the conical deficit of the spacetime. Specifically,
the conical deficits on the spatial two poles, $\theta_+=0$ (the
north pole ) and $\theta_-=\pi$ (the south pole ) are
\begin{equation}\label{deltapm}
\delta_\pm=2\pi\bigg[1-\frac{g(\theta_\pm)}{K}\bigg]\;.
\end{equation}
Such conical deficit corresponds to  the tension of the cosmic string~\cite{Appels:2016,Appels:2017}
\begin{equation} \label{mu}
\mu_\pm=\frac{\delta_\pm}{8\pi}=\frac{1}{4}-\frac{1\pm2mA+e^{2}A^{2}}{4K}\;.
\end{equation}
In order to avoid the occurrence of negative tension defects, clearly  we should require $0\leq\mu_+\leq\mu_{-}\leq1/4$.

To obtain the correct thermodynamics, we should first select an appropriate normalized time~\cite{Gibbons:2005}.
As far as  metric~(\ref{c-metric}) is concerned, such a normalization of the
time coordinate can be chosen such that it corresponds to the
"time" of an  observer comoving with the charged accelerating black hole~\cite{Anabalon:2018-2}, i.e., $\tau=\alpha{t}$ with
$\alpha=\sqrt{(1+A^2e^2)(1-A^2\ell^2-A^4\ell^2e^2)}$.  In following discussions, we will try to formulate the exact expressions of
the thermodynamic parameters for charged accelerating black holes in the frame of  normalized time $\tau$.
Now the thermodynamic mass  can be directly obtained via the conformal methods~\cite{Ashtekar:1999,Das:2000}
\begin{equation}
M=\big[1-(1+A^2e^2)A^2\ell^2\big]\frac{m}{K\alpha}\;,
\end{equation}
which is completely identified with the holographic mass~\cite{Anabalon:2018-2}.

The temperature $T$  can be found by using the
conventional Euclidean method associated with the normalized time,
\begin{equation}\label{T-H}
T=\frac{f'(r_+)}{4\alpha\pi}=\frac{m}{2\pi \alpha
r_{+}^{2}}-\frac{e^{2}}{2\pi \alpha
r_{+}^{3}}+\frac{A^{2}m}{2\alpha\pi}-\frac{A^{2}r_{+}}{2\alpha\pi}+\frac{r_{+}}{2\pi\alpha
\ell^{2}}\;,
\end{equation}
Here, note that in
order to have a well-defined temperature associated with the horizon
of the black hole, we assume that the acceleration is  small such
that the acceleration horizon can be negated by a negative
cosmological constant and only the black hole horizon
exists~\cite{Appels:2017}.

As usual, the  entropy $S$ of the accelerating black hole is one quarter
of the horizon area
\begin{equation}\label{S-H}
S=\frac{\pi r_{+}^{2}}{K(1-A^{2}r_{+}^{2})}\;.
\end{equation}
Meanwhile, the thermodynamic electric charge $Q$ of the black hole can be obtained by
\begin{equation}\label{Qch}
Q=\frac{1}{4\pi}\int_{\Omega=0}*dB=\frac{e}{K}\;,
\end{equation}
with the thermodynamic electrostatic potential
\begin{equation}\Phi=\frac{e}{{\alpha}r_+}\;\end{equation}
and the thermodynamic pressure  in extended black hole thermodynamics
\begin{equation} \label{PV}
P=-\frac{\Lambda}{8\pi}=\frac{3}{8\pi \ell^{2}}\;.\quad
\end{equation}

 We assume that all the relevant thermodynamic variables should obey the extended first law
\begin{equation}
\delta{M}=T\delta{S}+V\delta{P}+\Phi\delta{Q}-\lambda_+\delta\mu_+-\lambda_-\delta\mu_-\;.
\end{equation}
Then, according to the Smarr relation
\begin{equation}M=2TS-2PV+Q\Phi\;,\end{equation}
it is straightforward to verify
\begin{align}
&V=\frac{4\pi}{3 K\alpha}\Big[\frac{r_+^3}{(1-A^2r_+^2)^2}+A^2\ell^4m+A^4e^2\ell^4m\Big]\;,\nonumber\\
&\lambda_\pm=\frac{1}{\alpha}\Big[\frac{r_+}{1\pm{A}r_+}-\frac{m}{1+A^2e^2}\mp{A}\ell^2(1+A^2e^2)\Big]\;,\end{align}
where $\lambda_\pm$ are introduced as the thermodynamic lengths conjugate to the tensions $\mu_\pm$ \cite{Appels:2017,Anabalon:2018}.

For the case of vanishingly-small acceleration (i.e., $A\ell\ll1$
and $eA\ll1$), some proposed thermodynamic parameters can be
approximated as
 \begin{align}
 &M\approx\frac{{m}}{K}\Big[1-\frac{A^2}{2}\Big(e^2+\ell^2\Big)\Big]\;,\nonumber\\
 &V\approx\frac{4\pi}{3K}\Big[r_+^3+\frac{A^2}{2}\Big(2\ell^4m-e^2r_+^3+\ell^2r_+^3+4r_+^5\Big)\Big]\;,\nonumber\\
&\lambda_\pm\approx{r_+}-m\mp\Big(\ell^2+r_+^2\Big)A+\frac{A^2}{2}\Big[e^2\Big(3m-r_+\Big)+2r_+^3+\ell^2\Big(r_+-m\Big)\Big]\;.
 \end{align}

In order to gain a better understanding of the thermodynamic behavior,
let us recall the expression of the conical deficits in
Eq.~(\ref{deltapm}). As we see that the conical deficits in
this spacetime are unequal on different axes, so it is impossible to
choose a fixed $K$ to simultaneously regularize the metric at
both two poles.  Such kind of irregularity along an axis is just
precisely a definition of a conical singularity which encodes the
information about the presence of cosmic string. Actually, the
regularity of the metric at one pole demands
\begin{equation}
K_{\pm}=g(\theta_\pm)=1\pm2mA+e^{2}A^{2}\;.
\end{equation}
Here, we choose  $K=K_+=1+2m A+e^2A^2$  so that  the metric is just regular on the north pole (i.e., $\mu_+=0$),  leaving a conical deficit on the south pole, then the string tension on the south pole yields $\mu_-=mA/K\;.$

 Before further exploring  the thermodynamical behaviours  of charged accelerating  black holes,  a few comments concerning  the normalization factor $\alpha$ are in order. As stated previously,  the assumption of a "slowly accelerating black hole" ensures that the temperature of the black hole is well-defined,  the accelerating horizon is removed , and  all the thermodynamic variables in the first law are unambiguous. Explicitly, the slowly accelerating $C-$metric  means that the parameters of  the solution in metric~(\ref{c-metric}) satisfy $A\ell<1$  and $2Am<1$~\cite{Appels:2016,Appels:2017}. Furthermore, it is straightforward to verify that  $e^2A^2<1$ by considering the presence of the event horizon of the black hole, i.e, using $f(r_+)=0$ and $1>A r_+>0$. Thus, we have
\begin{equation}\label{alpha-series}
\frac{\alpha_0}{\alpha}=\frac{\sqrt{1-A^2\ell^2}}{\sqrt{(1+A^2e^2)(1-A^2\ell^2-A^4\ell^2e^2)}}=1-\frac{e^2A^2}{2}+\Big(\frac{3e^4}{8}+\frac{e^2\ell^2}{2}\Big)A^4+...\,,
\end{equation}
where  $\alpha_0£º=\sqrt{1-A^2\ell^2}$  denotes the normalization factor without charge.  As we can see form Eq.~(\ref{alpha-series}), the ratio $\alpha_0/\alpha$  approximates  to 1  for $A^2e^2\ll1$ and $A\ell<1$, and this just is the more slowly accelerating requirement. In addition, it is worth pointing out that, practically,  the value of the parameter $e$ (or $Q$) can not be  chosen arbitrarily large, and it should keep the corresponding charged black hole far away from the extremal case where the temperature vanishes.  Therefore, the difference between $\alpha_0$ and $\alpha$ usually is tiny if the slowly accelerating assumption ($A^2e^2\ll1$ and $A\ell<1$) is strictly applied.

\begin{figure}[ht]
\centering{\subfigure[]{\label{M-lambda11}
\includegraphics[width=0.54\textwidth]{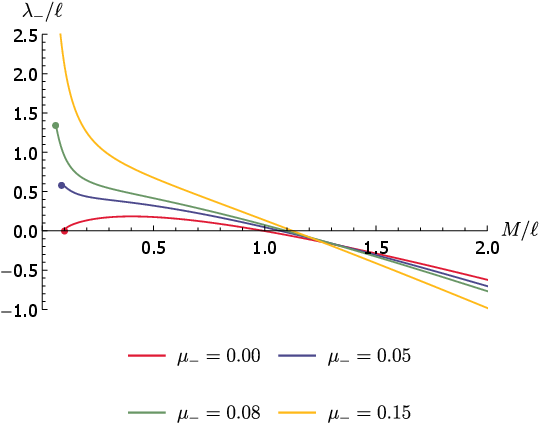}}\subfigure[]{\label{M-lambda22}
\includegraphics[width=0.54\textwidth]{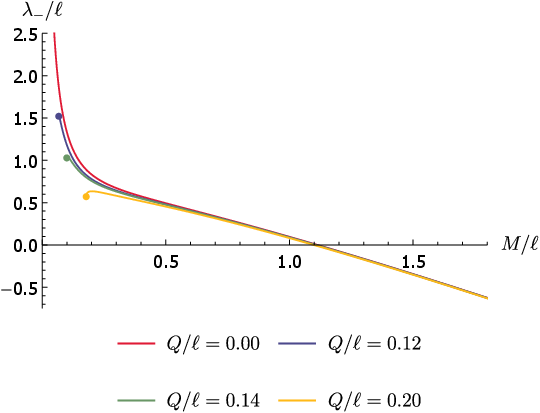}}}
\caption{The thermodynamic length of the south pole is plotted as a
function of the renormalized thermodynamic mass $M$ for (a) the
varying  tension with a fixed $Q/\ell=0.10$, and (b) the varying
$Q/\ell$ with a fixed tension $\mu_-=0.10$. All the solid circle
points represent the corresponding extremal black
holes.}\label{M-lambda}
\end{figure}

 Let us return to the  thermodynamic variables. We can see, from figure~\ref{M-lambda}, that when the renormalized thermodynamic mass $M$ is comparable to AdS radius $\ell$,
 the thermodynamic length $\lambda_-$ becomes negative. And  the thermodynamic length is not always a monotonic decreasing
 function of $M$ in the region of $M/\ell<1$, and the functional behavior  depends on the concrete values of  $\mu_-$ and $Q$. Furthermore, when
 $M$ becomes very large  for a fixed tension $\mu_-$, the influence of the charge variation on the thermodynamic length almost disappears.
 Here, notice that if $M/\ell$ is smaller than that of the extremal black hole, the temperature
becomes negative, which is  usually considered as unphysical.
Moreover, it should be pointed out that as $A\ell$ is drawing near
one,  the presence of the acceleration horizon makes the thermodynamic
temperature ambiguous~\cite{Appels:2016,Appels:2017,Anabalon:2018}.
So, the yellow curve ($\mu_-=0.15$) in figure~\ref{M-lambda11} does
not continue to arbitrarily small $M/\ell$ (i.e.,
$M/\ell\gtrsim0.0232$ or equivalently $A\ell\lesssim0.9783$) in
order  for the thermodynamic temperature of accelerating black holes
to make sense. Neither does the red curve ($Q/\ell=0$) in
figure~\ref{M-lambda22} with $M/\ell\gtrsim0.0123$ (or equivalently
$A\ell\lesssim0.9925$).

\begin{figure}[ht]
\centering{\subfigure[]{\label{T-Mplot11}
\includegraphics[width=0.54\textwidth]{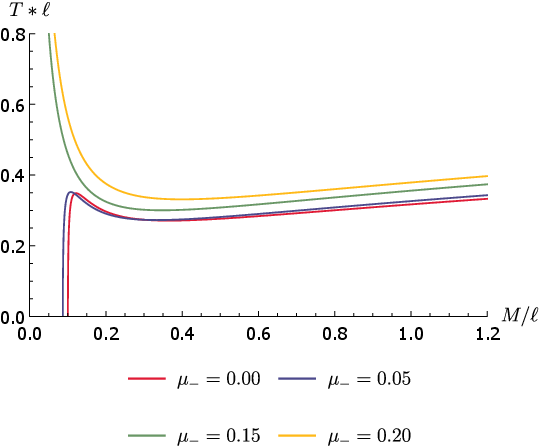}}\subfigure[]{\label{T-Mplot22}
\includegraphics[width=0.54\textwidth]{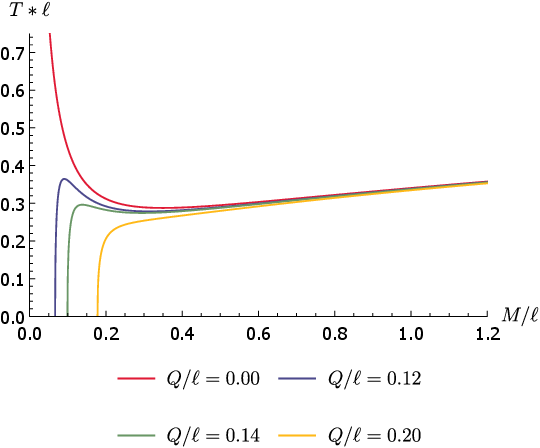}}}
\caption{ A plot of the temperature  as a function of the thermodynamic mass for varying string tension $\mu_-$ with fixed $Q/\ell=0.10$  in (a), and for varying $Q/\ell$  with  fixed tension $\mu_-=0.10$ in (b).}\label{T-Mplot}
\end{figure}

A representative plot of the temperature as a function of the mass, $M$ , for the accelerating charged black hole  is illustrated in figure~\ref{T-Mplot}.
As we can see from  figure~\ref{T-Mplot11}, if the string tension $\mu_-$ is small, the temperature can approach zero as the mass decreases, signaling an extremal black hole.  However, as
 $\mu_-$  increases, there seems to be a critical value of  $\mu_-$  above which the extremality never occurs and the corresponding  temperature has a local positive minimum value.
 It is difficult to obtain an  analytic expression of the critical point. However, numerical calculations indicate that  the value is approximately $0.10$ with very high precision, which happens to be the
 value of $Q/\ell$. In this situation, it should be emphasized  that $M/\ell$  cannot go arbitrarily small in order to avoid the occurrence of the accelerating horizon. For example, for  the case of $\mu_-=0.15$,  $M/\ell\gtrsim0.0232$ is required  (or equivalently $A\ell\lesssim0.9783$) as aforesaid, while for $\mu_-=0.20$, $M/\ell\gtrsim0.0536$ (or equivalently $A\ell\lesssim0.9534$).
Similarly, for a varying $Q/\ell$, as shown in figure~\ref{T-Mplot22}, the increasing $Q/\ell$ will make the charged accelerating black hole approach the extremal limit at a small $M/\ell$. If $Q/\ell$ is large enough, then temperature becomes a monotonic function of the mass and the extremality no longer happens at large enough $M/\ell$ .

\begin{figure}[ht]
\centering{\subfigure[]{\label{F-Tplot11}
\includegraphics[width=0.54\textwidth]{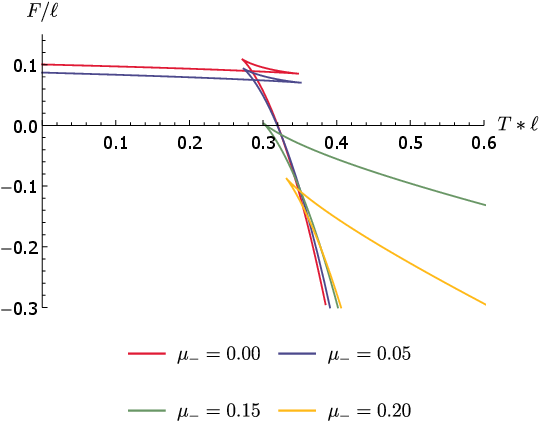}}\subfigure[]{\label{F-Tplot22}
\includegraphics[width=0.54\textwidth]{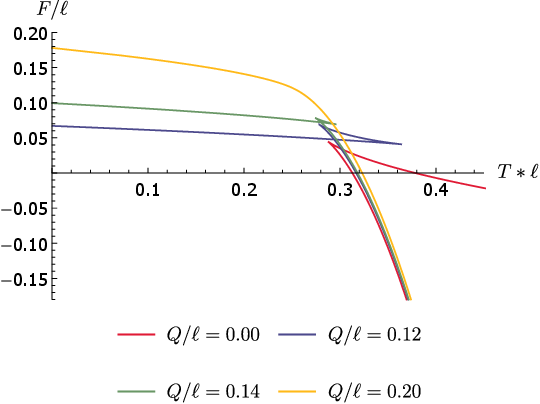}}}
\caption{ The free energy is plotted  as a function of temperature for varying string tension $\mu_-$ with fixed $Q/\ell=0.10$  in (a), and varying $Q/\ell$  with  fixed $\mu_-=0.10$ in (b).}\label{F-Tplot}
\end{figure}

In figure~\ref{F-Tplot}, we have displayed the variation of the free
energy $F=M-TS$  with the temperature  for varying tension and
charge in the canonical ensemble. For the case of a fixed $Q/\ell=0.10$,
as the tension $\mu_-$ is increased, the swallowtail style will be
suppressed, leaving only two wings, which means that the increasing
tension renders the charged accelerating black holes far away from
extremality, and a spontaneous phase transition from a small black
hole to a large black hole will occur. For a fixed tension, the
prominent swallowtail exists in the region where $Q/\ell $ becomes
comparable to $\mu_-$. As $Q/\ell$ is gradually increased or
decreased, the swallowtail will disappear. Especially for large
enough $Q/\ell$, only a single branch is left at last, which
means that there is a critical point for $Q/\ell$ at which the
intermediate regime with negative specific heat for the charged accelerating
black hole just disappears.

\section{Benchmarking Black hole heat engines and efficiency}
\label{III}
With the first law of charged  accelerating  AdS black holes derived, now we turn our attention to the features of the corresponding  holographic heat engine in the benchmarking scheme.
It is well-known that the efficiency for a valid heat cycle is defined by
\begin{align}
\eta=1-\frac{Q_C}{Q_H}\;,
\end{align}
where $Q_C$ denotes a  net output heat flow in one cycle,  and $Q_H$
represents a net input heat flow. In  the benchmarking
scheme~\cite{Johnson:2016-1,Hennigar:2017-1}, a circular or
elliptical cycle has been suggested  to allow for
cross-comparison of the efficiencies of holographic heat engines
with different black holes as the working substances. In the $P-V$
plane, a general circle  cycle can be  written as
\begin{align}\label{PV-2}
P(\theta)=P_{0}+R\sin\theta\;, V(\theta)=V_{0}+R\cos\theta\;,
\end{align}
where the circle is centered at $(V_0,P_0)$ with a radius of $R$. Note that the relation  $R<P_0$ and $R<V_0$
should hold for a physical heat cycle.

As we have shown in the proceeding  section, all the thermodynamic
variables in the first law are  certain combinations  of the black hole
parameters and the specific heat capacity at constant volume is
nonzero (i.e., $C_V\neq0$). Therefore, it is difficult to obtain the
analytic expression for the efficiency of benchmarking black hole
heat engines and a numerical integration of $TdS$ will be needed.

In order to better understand how the holographic heat engine of  the accelerating charged black holes  behaves, let us first review the influence of a charge on the  temperature of the black holes. According to Eq.~(\ref{T-H}), it is easy to see that the contribution of a charge to the temperature is negative. Therefore, there exists an  upper-bound  of $Q$ for the black hole thermodynamics.  If  $Q$ approaches the upper-bound, the temperature  becomes zero and the black hole becomes extremal. Since a valid cycle contour is forbidden to  cross the area of non-positive temperature, a zero isothermal curve cannot be tangent  to or intersecting with the  circular cycle in the $P-V$ plane.

\begin{figure}[!ht]
\centering{\subfigure[]{\label{Qmax11}
\includegraphics[width=0.49\textwidth]{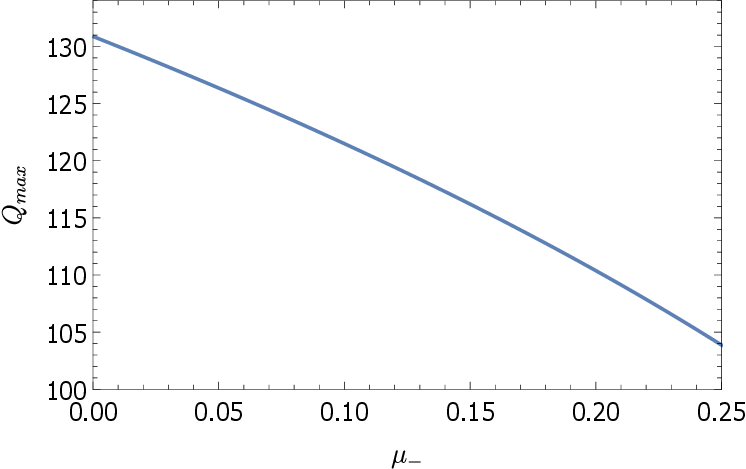}}
\subfigure[]{\label{Qmax22}\includegraphics[width=0.49\textwidth]{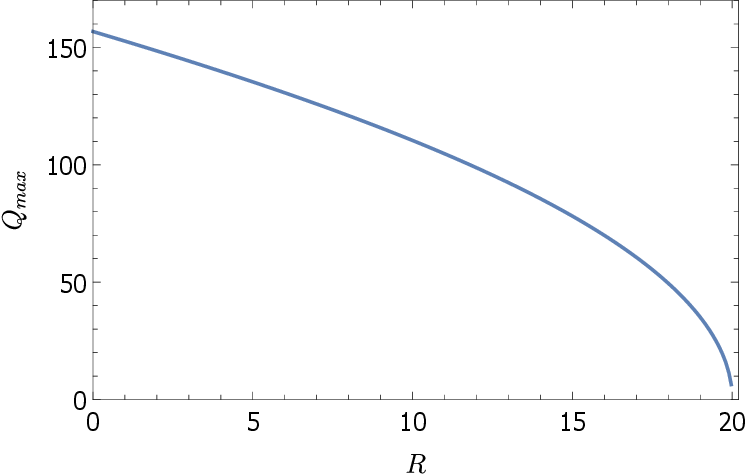}}}
\caption{Assuming the benchmarking circle is  at $(V_{0},P_{0})=(100,20)$ in the $P-V$ plane, the upper-bound of charge $Q_{max}$ is plotted as a function of (a)$\mu_-$ for fixed $R=10$ and (b) $R$ for fixed $\mu_-=0.20$.   }\label{Qmax}
\end{figure}

For a benchmarking circle in the $P-V$ plane, there must be a point of minim temperature on the circle. If we suppose an extremal black hole at the point of minim temperature on the cycle, then the corresponding value of $Q$ of the extremal black hole can be considered as an extremal upper-bound of charge for the benchmarking circle.  This upper-bound  is denoted by  $Q_{max}$ in our following discussions.  It is not difficult to deduce that if the thermodynamic charge is larger than $Q_{max}$,  negative temperature occurs  and the cycle process stops. Therefore, a benchmarking heat engine of black holes with charge can make physical sense only when $Q<Q_{max}$.   In figure~\ref{Qmax}, we have displayed how $Q_{max}$ changes as a function of  $\mu_-$ and $R$.
 We can see that the upper-bound $Q_{max}$ is a decreasing function of  $\mu_-$ and $R$. Here, it should be emphasized that the symbol $Q_{max}$ or $Q$ in the following plots is not in units of $\ell$ but in that of charge.

\begin{figure}[hbp]
\centering{
\includegraphics[width=0.7\textwidth]{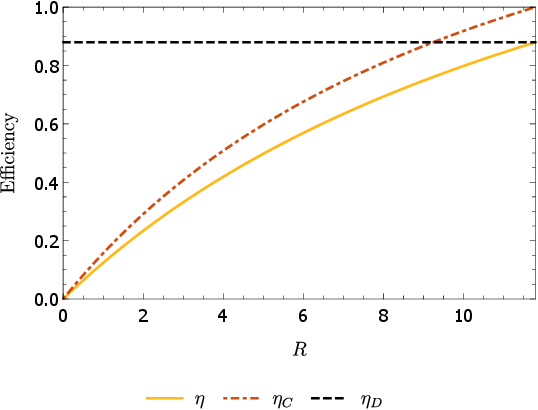}}
\caption{ The efficiency of a heat engine of charged accelerating black holes  is plotted as a function of  $R$ with fixed $Q=100$ and $\mu_-=0.20$,  where $\eta_C$ denotes Carnot efficiency and $\eta_{D}=2\pi/(\pi+4)$ means the upper-bound value of efficiency. We have assumed the centre of the cycle satisfies $(V_0,P_0)=(100,20)$. Here, for a valid heat engine cycle, $R$ should be smaller than $11.7955$  since the upper-bond $Q_{max}=100$ for $R\approx11.7955$ and $\mu_-=0.20$.}\label{eta-R-fixcmQ}
\end{figure}

Let us return to the efficiency in the benchmarking circular cycle in Eq.~(\ref{PV-2}), which will be influenced by  the thermodynamic variables $Q$,  $\mu_-$ and the circle radius $R$.
We first present the behavior of the efficiency with the increasing circle radius $R$ for fixed $Q$ and $\mu_-$ in figure~\ref{eta-R-fixcmQ}.
Here, the Carnot efficiency is classically defined by $\eta_C=1-T_C/T_H$ with  maximum temperature $T_H$ and minim temperature $T_C$ in the entire cycle process, and $\eta_D={2\pi}/(\pi+4)$ is proposed as the universal upper-bound for the efficiencies of benchmarking holographic heat engines on circular cycles in Ref.~\cite{Hennigar:2017-1}.
As we can see,  the efficiency of the  benchmarking black hole heat engine $\eta$ increases  with  the increase of $R$,  and the corresponding Carnot efficiency $\eta_C$ is always larger than the efficiency $\eta$. Besides these, $\eta$ can approximate to the upper-bound $\eta_{D}$ so long as $R$ draws near $11.7955$.  A comment is in order here.  According to Eq.~(\ref{T-H}) and Eq.~(\ref{PV-2}), it is not difficult to derive that $Q_{max}$,  the upper-bound of charge, will approximate to  $100$ for $\mu_-=0.20$ and $R=11.7955$.  If $R$  goes beyond $11.7955$, the $Q_{max}$ will become smaller than $100$ (see figure~\ref{Qmax22}), then the circular cycle for $Q=100$ will cross the region of negative temperature, which makes no sense in physics. Therefore, in the setting of $Q=100$,  $\mu_-=0.20$ and $(V_0,P_0)=(100,20)$,   $R$ cannot go beyond $11.7955$,  and the efficiency will approximate to the upper-bound $2\pi/(\pi+4)$ in the limit of $R$ approaching $11.7955$.

\begin{figure}[ht]
\centering{\subfigure[]{\label{effi-r11}
\includegraphics[width=0.52\textwidth]{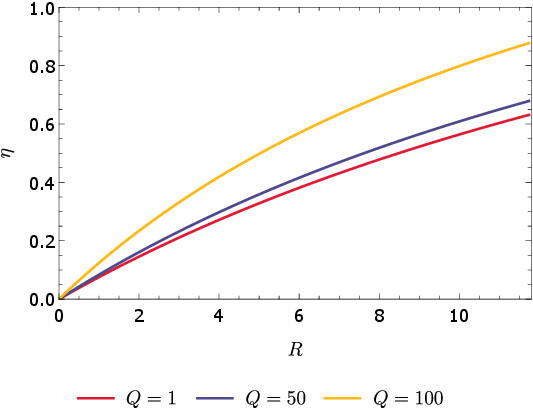}}\subfigure[]{\label{effi-r22}
\includegraphics[width=0.52\textwidth]{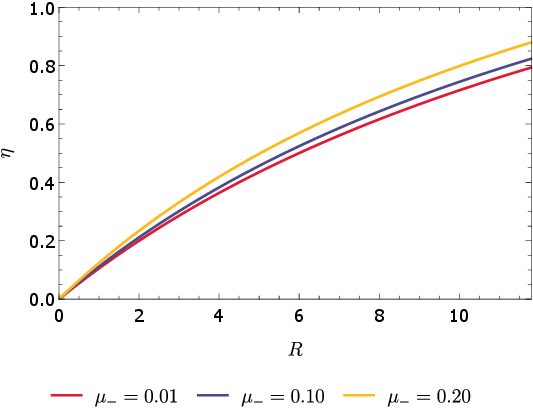}}}
\caption{The efficiency of the benchmarking heat engine as a function of the radius $R$ for the circular cycle is centered at $(V_0,P_0)=(100,20)$.   (a)Thermodynamic charge $Q=\{1,50,100\}$ with fixed cosmic string tension $\mu_-=0.20$. (b) The tension $\mu_-=\{0.01,0.10,0.20\}$ with fixed  $Q=100$.    }\label{qandmu-effi-R}
\end{figure}

In figure~\ref{qandmu-effi-R},  the efficiencies are plotted  as a function of the circle radius $R$   for fixed cosmic string tension and fixed charge respectively. It is easy to find that the efficiency is still an increasing function of $R$.  For a fixed $R$, the larger the $Q$ or the larger the $\mu_-$, the higher  the efficiency.

\begin{figure}[!ht]
\centering{
\includegraphics[width=0.65\textwidth]{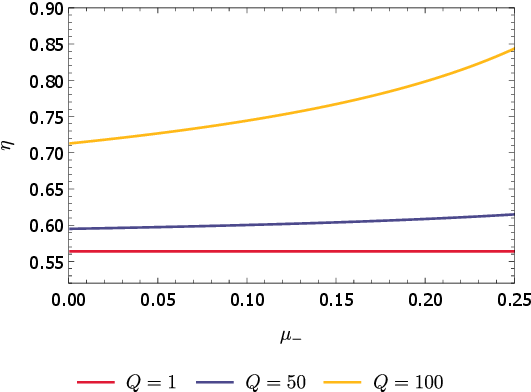}}
\caption{ The efficiency of the holographic heat engine as a function of cosmic string tension $\mu_-$ with  $Q=\{1,50,100\}$. The center of the circle is  at $(V_{0},P_{0})=(100,20)$ with a fixed radius $R=10$.  }\label{mu-q}
\end{figure}

In figure~\ref{mu-q}, we  give a representative plot of the efficiency as a function of the cosmic string tension  $\mu_-$  for some different values of charge $Q$,  which illustrates  how the variation of charge influences the curved contour describing the behavior of the efficiency with increasing $\mu_-$ for a fixed benchmarking circle.  One can see  that when the value of charge $Q$ is small a variation of $\mu_-$ does not cause significant changes in the efficiency as one would expect.  However, when the value of $Q$ is large, the efficiency becomes an obvious increasing function of the cosmic string tension $\mu_-$ in contrast to the uncharged case~\cite{Zhang:2018}. Therefore, the presence of a charge may significantly increase the efficiency of the benchmarking heat engine with slowly accelerating AdS black holes as working substances.

\begin{figure}[ht]
\centering{
\includegraphics[width=0.7\textwidth]{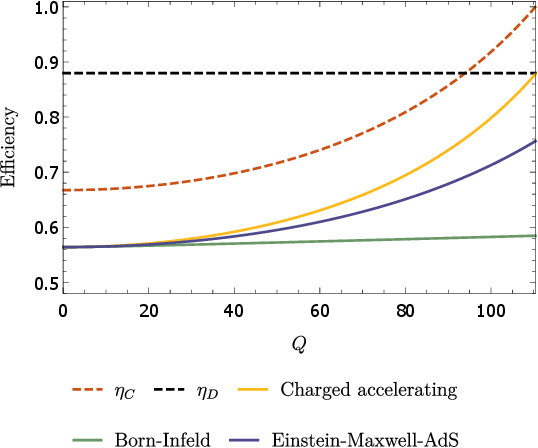}}
\caption{ The efficiencies vs electric charge $Q$   with the center of benchmarking cycle localized at $(V_{0},P_{0})=(100,20)$ and $R=10$.  The three  solid curves respectively represent  the efficiencies of three  different families of  black holes as the working material, including charged accelerating  AdS ($\mu_-=0.20$), Born-Infeld ($\beta=1$) and Einstein-Maxwell-AdS  black holes. Note that the extreme charge  for charged accelerating AdS black holes in this benchmarking cycle reads $Q_{max}\approx110.3813$, of which the efficiency will draw near $\eta_D={2\pi}/({4+\pi})$.  Here, $\eta_C$ denotes the Carnot efficiency of charged accelerating  black hole in this circular cycle.}\label{effi-Q-com}
\end{figure}

Now, we cross-compare  the efficiency of different black hole heat engines in the benchmarking scheme.  The black hole  heat engines we choose are those of the slowly accelerating charged AdS black holes we just studied, the Einstein-Maxwell-AdS black holes (Reissner-Nordstrom-like black hole) and the Born-Infeld black holes. We plot the efficiency of the benchmarking heat  engine as a function of charge $Q$  with different black holes as working substances  in  figure~\ref{effi-Q-com}. It is worth noting that both  Einstein-Maxwell-AdS black hole  and Born-Infeld black hole have a vanishing specific heat at constant volume(i.e., $C_V=0$, the detailed thermodynamic quantities can be found in Ref.~\cite{Johnson:2016-1,Chamblin:1999,Cai:2004,Dey:2004}), and the  holographic heat engine efficiency has been studied in  Ref.~\cite{Hennigar:2017-1,Chakraborty:2016}.

 Figure~\ref{effi-Q-com} reveals that the existence of charge in general increases the work efficiency to some extent. More interestingly, one finds that with acceleration, the holographic heat engine efficiency is usually larger than that without, and moreover  it increases more rapidly as the charge grows.

\section{Conclusion}
\label{IV}
We have explored the thermodynamics of charged accelerating  AdS black holes described by $C-$metric. In order to obtain the consistent thermodynamic depiction for the charged accelerating black holes that have an uncommon asymptotical behaviour, the normalization of the timelike Killing vector is fixed  by holographic methods, i.e., $$\tau =\sqrt{(1+A^2e^2)(1-A^2\ell^2-A^4\ell^2e^2)}\;t\;,$$  and such a choice makes all the corresponding thermodynamic parameters be analogous  to those of an observer comoving with the accelerating charged object.
Through some algebraic manipulations, we find that
 the exact expression of the  thermodynamic mass
in the consistent first law is complicatedly intertwined with the parameters $e,A,\ell$, so do the  other thermodynamic quantities, such as, the volume and the thermodynamic length. With the first law, the properties of the  thermodynamic length and
the possible phase behaviour of the charged accelerating black holes are also
discussed in a standard thermodynamic analysis. We find that the increasing tension renders  the  charged accelerating black hole away from extremality, the phase transition to large black holes occurs, while increasing the charge, the intermediate regime with negative specific heat will  gradually disappear.

Moreover,  we have examined the features of the  heat engines with charged accelerating  black holes as the working substances in the benchmarking scheme of holographic heat engines. Due to the fact that the corresponding  thermodynamic variables in the first law are non-linear combinations of the black hole parameters $r_+$, $ m$, $e$, $\ell$ and $A$, the numerical estimation is used  to demonstrate the behavior of the efficiency of the holographic heat engines.
We found that the efficiency of the holographic  heat  engine  can be influenced by the cosmic string tension, the size of circular cycle and the thermodynamic charge. When the cosmic string tension is fixed, the efficiency of the holographic heat engine is increased by the existence of charge. For a fixed size of the circular cycle,  the existence of charge may significantly increase the efficiency of the holographic heat engine as the string tension grows. A cross-comparison of the holographic heat engines with slowly accelerating charged AdS black holes and the  Einstein-Maxwell-AdS black holes in a fixed benchmarking cycle shows that  the presence of acceleration also increases the efficiency, and moreover, it makes the efficiency more sensitive to a varying charge $Q$.

\begin{acknowledgments}
 This work was supported by the National Natural Science Foundation of China under Grants  No. 11435006 and No.11690034.
\end{acknowledgments}


\end{document}